\documentstyle[epsfig]{l-aa}

\newcommand\yr{$\rm yr^{-1}$}
\def\ltsima{$\; \buildrel < \over \sim \;$}
\def\simlt{\lower.5ex\hbox{\ltsima}}
\def\gtsima{$\; \buildrel > \over \sim \;$}
\def\simgt{\lower.5ex\hbox{\gtsima}}
 
\begin{document}
   



   \thesaurus{13.25.2 -- 11.03.4}

   \title{A BeppoSAX observation of the cooling flow cluster Abell 2029}

   \author{Silvano Molendi
          \inst{1}
   \and Sabrina De Grandi
	  \inst{2}
}

   \offprints{S.Molendi, silvano@ifctr.mi.cnr.it}

   \institute{Istituto di Fisica Cosmica ``G.Occhialini'', Via Bassini 15,
              I--20133 Milano, Italy
   \and Osservatorio Astronomico di Brera, Via Bianchi 46,
        I-23807 Merate (LC), Italy
}

   \date{Received / Accepted }

   \maketitle

\markboth{S. Molendi and S. De Grandi}{BeppoSAX observation of Abell 2029}

\begin{abstract}
We present results from a BeppoSAX observation of the cooling flow cluster 
Abell 2029.
The broad band spectrum (2-35 keV) of the cluster, 
when fitted with a model including a cooling
flow component and a one temperature thermal component, yields
a temperature of 8.3$\pm$0.2 keV and a metal abundance of 
0.46$\pm$0.03 in solar units, for the latter.  
No evidence of a hard X-ray excess is found in the PDS spectrum. 
By performing a spatially resolved spectral analysis we find
that the projected temperature and abundance drop with increasing radius, 
going from $\sim$ 8 keV and 0.5 (solar units) at the cluster core to 
$\sim$ 5 keV and  0.2 (solar units) at about 1.2 Mpc.


\end{abstract}


\section {Introduction}

Abell 2029 (hereafter A2029)
is a rich, nearby  (z$=$ 0.0766), X-ray luminous, cluster of galaxies.
In the optical band, Oegerle et al. (1995), by analyzing the velocity 
dispersion of a large  number of galaxies, do not find any strong evidence
of substructure in A2029. X-ray observations (e.g. Slezak, Durret \&
Gerbal 1994; Buote \& Canizares 1996) provide clear evidence that A2029
is a regular cluster. 
Various authors, either by deprojection analysis of ROSAT data (e.g. 
Sarazin, O'Connell \& McNamara 1992; Peres et al. 1998)
or through spectral analysis of ASCA and ROSAT data (e.g. Sarazin, Wise 
\& Markevitch 1998, hereafter S98), have measured a substantial cooling 
flow in the core of A2029.  
David et al. (1993), using Einstein MPC data, report a global temperature
of 7.8$^{+0.8}_{-0.7}$ keV.  
S98, from the analysis of ASCA data, find evidence of a temperature 
gradient.
The projected temperature is found to decrease from $\sim$ 9 keV to 
$\sim$ 6 keV when going from the cluster core out to  $\sim$ 1.6 Mpc.
The temperature map of A2029, presented by S98,
is consistent with an azimuthally symmetric temperature pattern.
Irwin, Bregman \& Evrard (1999), who have used ROSAT PSPC data to search 
for temperature gradients in a sample of galaxy clusters including A2029,
in agreement with  S98, find evidence of a 
radial temperature gradient in A2029.
Using ASCA data Allen \& Fabian (1998) measure an average metal abundance 
of 0.46$\pm$0.03. 
S98, again using ASCA data, do not find compelling evidence 
of an abundance gradient in  A2029.

In this Letter we report a recent BeppoSAX observation of A2029.
We use our data to perform an independent measurement of the 
temperature profile and two-dimensional map of A2029.
We also present the abundance profile and the first abundance map 
of A2029. 
The outline of the Letter is as follows.
In section 2 we give some information on the BeppoSAX observation
of A2029 and on the data preparation.
In section 3 we present the analysis of the broad band spectrum
(2-35 keV).
In section 4  we present spatially resolved measurements
of the temperature and metal abundance.
In section 5 we discuss our results and compare them to previous
findings.
Throughout this Letter we assume H$_{o}$=50 km s$^{-1}$Mpc$^{-1}$
and q$_{o}$=0.5.
 
\section {Observation and Data Preparation}
The cluster A2029  was observed by the BeppoSAX 
satellite (Boella et al. 1997a) 
between the 4$^{th}$ and the 5$^{th}$ of February 1998.
We will discuss here data from two of the instruments onboard BeppoSAX:
the MECS and the PDS.
The MECS (Boella et al. 1997b)
is presently composed 
of two units (after the failure of a third one), working in the 1--10 keV 
energy range. At 6~keV, the energy resolution is  $\sim$8\%  and the 
angular resolution 
is $\sim$0.7$^{\prime}$ (FWHM). The PDS instrument (Frontera et al. 1997), 
is a passively collimated detector (about 1.5$\times$1.5 degrees f.o.v.), 
working in the 13--200 keV energy range.
Standard reduction procedures and screening criteria have been
adopted to produce linearized and equalized event files.
Both MECS and PDS data preparation and linearization was performed using 
the {\sc Saxdas} package.
The effective exposure time of the observation was
 4.2$\times$10$^4$ s (MECS) and 1.8$\times$10$^4$ s (PDS). 
The observed countrate for A2029 was 0.812$\pm$0.004 cts/s for the 2 
MECS units and 0.19$\pm$0.04 cts/s for the PDS instrument.

All MECS spectra discussed in this Letter have been background 
subtracted using spectra extracted from blank sky event files in the same 
region of the detector as the source. The energy range considered for
spectral fitting is always 2-10 keV.
All spectral fits have been performed using XSPEC Ver. 10.00.
 Quoted confidence 
intervals are 68\% for 1 interesting parameter (i.e. $\Delta \chi^2 =1$), 
unless otherwise stated.

\section{Broad Band Spectroscopy} 

We have extracted a MECS spectrum,  
from a circular region of 8$^{\prime}$ radius
(0.95 Mpc), centered on the  
emission peak. From the ROSAT PSPC radial profile,
we estimate that about 90\% of the total cluster emission falls
within this radius.   
The PDS background-subtracted
spectrum has been produced by subtraction
of the ``off-source'' from the ``on-source'' spectrum.
As in Molendi et al. (1999) (hereafter M99), a numerical relative 
normalization factor among MECS and PDS spectra has been included to 
account for: the fact that the MECS includes emission out to about 1 Mpc from 
the X-ray peak, while the PDS field of view covers the whole cluster;
the mismatch in the absolute flux calibration of the MECS and PDS 
response matrices; the vignetting in the PDS instrument.
The estimated normalization factor is  0.76. In the fitting procedure we 
allow this factor to vary within 15\% from the above value to account for 
the uncertainty in this parameter. 
The spectra from the two instruments have been fitted 
with a one temperature thermal emission component plus a cooling 
flow component (MEKAL and MKCFLOW codes in the XSPEC 
package), absorbed by a galactic line of sight equivalent hydrogen 
column density, $N_H$, of 3.05$\times 10^{20}$ cm$^{-2}$
(Dickey \& Lockman 1990). 
All parameters of the cooling flow component were fixed,
as the energy range we use for spectral fitting (2-35 keV) is not 
particularly sensitive to this component.
More specifically, the minimum temperature 
was fixed at 0.1 keV, the maximum 
temperature, and the metal abundance were set to be equal to 
the temperature  and the metal abundance
of the MEKAL component, the deposited mass, $\dot M$,  
was fixed at the value of 363$M_\odot$\yr derived by 
S98 when fitting ROSAT PSPC and ASCA GIS data.  
The model yields an acceptable fit to the data, $\chi^2 =$ 160.0 
for  167 d.o.f. The best fitting values for the temperature
and the metal abundance are  respectively, 8.3$\pm$0.2 keV
and  0.46$\pm$0.03, solar units.
By assuming a value of $\dot M = 556 M_\odot$\yr, equal to the 
one derived by Peres et. al (1998) by deprojecting the ROSAT PSPC 
surface brightness profile, we obtain a fit of similar quality
$\chi^2 =162.3 $  for 167 d.o.f. and derive a slightly higher value
for the temperature  8.6$\pm$0.2 keV and a similar 
value for the abundance, 0.47$\pm$0.02. 

\vspace{-0.20in}
\begin{figure}
\centerline{
      \hbox{
      \psfig{figure=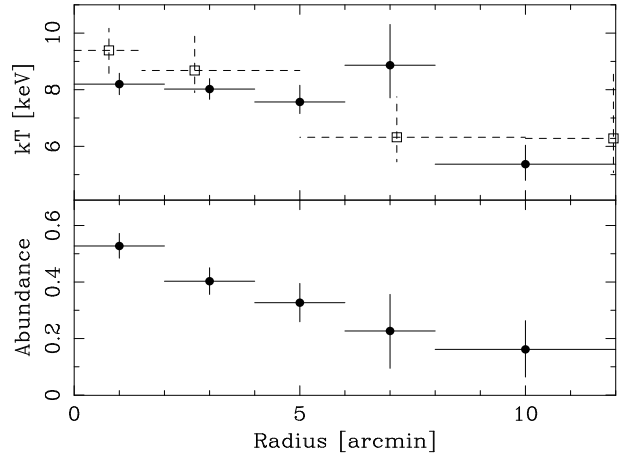,height=6.5cm,width=9cm,angle=-90}
}}
\vspace{-0.20in}
\caption{
{\bf Top Panel}: projected radial temperature profile.
The open squares and filled circles indicate respectively the ASCA 
measurements, obtained by S98, and our own BeppoSAX measurements.
All the uncertainties on the
temperature measurements are at the 68$\%$ confidence level (we have
converted the 90$\%$ confidence errors reported in figure 4 of
S98 into 68$\%$ confidence errors by dividing
them by 1.65, for further details see Markevitch \& Vikhlinin 1997).  
{\bf Bottom Panel}: projected radial abundance profile from BeppoSAX MECS
data. 
}
\label{figmg7}
\end{figure}
\vspace{-0.15in}
\section{Spatially Resolved Spectral Analysis} 
The spectral distorsions introduced by the energy dependent 
MECS PSF,  when performing spatially resolved spectral 
analysis, have been taken into account using the 
method described in M99 and references
therein.


We have accumulated spectra from 5 annular regions 
centered on the X-ray emission peak, 
with inner and outer radii of 0$^{\prime}$-2$^{\prime}$, 
2$^{\prime}$-4$^{\prime}$, 4$^{\prime}$-6$^{\prime}$, 
6$^{\prime}$-8$^{\prime}$ and 8$^{\prime}$-12$^{\prime}$.
A correction for the absorption caused by the strongback supporting
the detector window has been applied for  the  
8$^{\prime}$-12$^{\prime}$ annulus, where the annular part of 
the strongback is contained. For the 6$^{\prime}$-8$^{\prime}$  
region, where the strongback
covers only a small fraction of the available area, we have chosen
to exclude the regions shadowed by the strongback.
 
We have fitted each spectrum, except the one extracted from
the innermost region, with a MEKAL model absorbed 
by the galactic N$_H$, of 3.05$\times 10^{20}$ cm$^{-2}$. 
In the spectrum from the 0$^{\prime}$-2$^{\prime}$ region
we have included a cooling flow component, the parameters of 
this component have all been fixed, as in the fitting of the broad band
spectrum (see section 3).
The temperature and abundance we derive for the innermost region 
are respectively 8.2 $\pm$ 0.4 keV and 0.53 $\pm$ 0.04, solar units, 
if we assume the mass deposition reported by S98, 
$\dot M = 363 M_\odot$\yr,  and  9.0 $\pm$ 0.5 keV and 0.55 
$\pm$ 0.05, solar units, if we assume the mass deposition reported 
by Peres et. al (1998), $\dot M = 556 M_\odot$\yr. 
In figure 1 we show the temperature and abundance 
profiles obtained from the spectral fits, the values reported 
for the innermost annulus are those obtained by fixing the 
mass deposition to $\dot M = 363 M_\odot$\yr.  
Our measurements are practically unaltered if we excise from 
the accumulated spectra the emission of 
2 pointlike sources, clearly recognizable in the ROSAT PSPC image of A2029
(see figure 1 of S98).
By fitting the temperature and abundance profiles with a 
constant we derive the following average values: 
$ 7.7\pm$0.2 keV and 0.41$\pm$0.03, solar units. 
A constant does not provide an acceptable fit to the temperature profile.
Using the $\chi^2$ statistics we find: $\chi^2 =$15.2 for 5 d.o.f., 
corresponding to a probability of  0.009 for the observed distribution to 
be drawn from a constant parent distribution.
A linear profile of the type, kT = a $ + $ b~r, where kT is in keV and
r in arcminutes, provides a much better fit, $\chi^2 =$ 3.5 
for 4 d.o.f. The best fitting values for the parameters are 
a$ = 8.7 \pm 0.4$ keV, b$= - ~ 0.28 \pm 0.08$ keV~arcmin$^{-1}$.
The improvement is found to be statistically significant at more 
than the 97.5\% level according to the F-test.
As for the temperature, a constant does not provide an acceptable
fit to the abundance profile, $\chi^2 =$16.4 for 4 d.o.f. (Prob.$=$0.002). 
Interestingly, a linear profile of the type, Ab = a $+$ b~r, where r is 
in arcminutes and Ab is in solar units, provides a significantly better fit, 
$\chi^2 =$0.6 for 3 d.o.f. According to the F-test, the probability that
the improvement in the fit might be associated to the reduction in the
d.o.f. is $<$0.001. The best fitting values for the parameters are 
a$=0.55\pm 0.04$ solar units, b$=- 0.043\pm 0.011$ solar units~arcmin$^{-1}$.
   


\vspace{-0.15in}
\begin{figure}
\centerline{
      \hbox{
      \psfig{figure=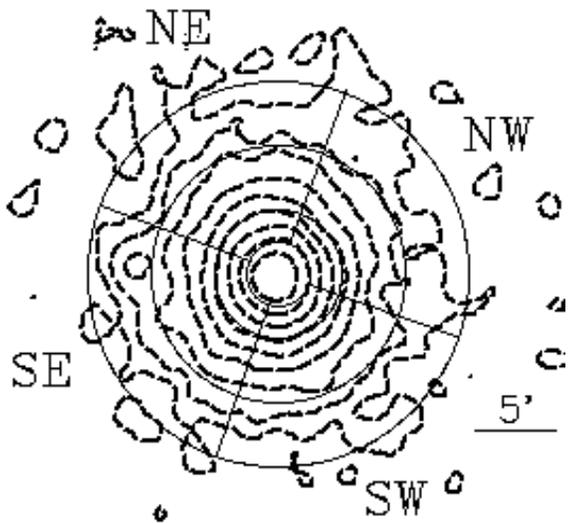,height=7.0cm,width=7.5cm,angle=0}
}}
\vspace{-0.15in}
\caption{
MECS image of A2029. Logarithmic contour levels are indicated
by the dashed lines. The solid lines show how the cluster
has been divided to obtain temperature and abundance maps.
}
\label{figmg7}
\end{figure}
\vspace{-0.15in}
We have divided A2029 into 4 sectors: NW, SW, 
SE and NE. 
Each sector has been divided into 3 annuli with bounding radii,
 2$^{\prime}$-4$^{\prime}$, 4$^{\prime}$-8$^{\prime}$ and 
8$^{\prime}$-12$^{\prime}$. In figure 2 
we  show the MECS image with the sectors overlaid.
A correction for the absorption caused by the strongback supporting
the detector window has been applied for  the sectors  
of the 8$^{\prime}$-12$^{\prime}$ annulus.
We have fitted each spectrum with a MEKAL model absorbed 
by the galactic N$_H$. 
Our temperature and abundance measurements are practically unaltered if 
we excise from the spectra the emission of 
the 2 pointlike sources visible in the ROSAT PSPC image of A2029
(see figure 1 of S98).

\vspace{-0.05in}
\begin{figure}
\vspace{-0.20in}
\centerline{
      \hbox{
      \psfig{figure=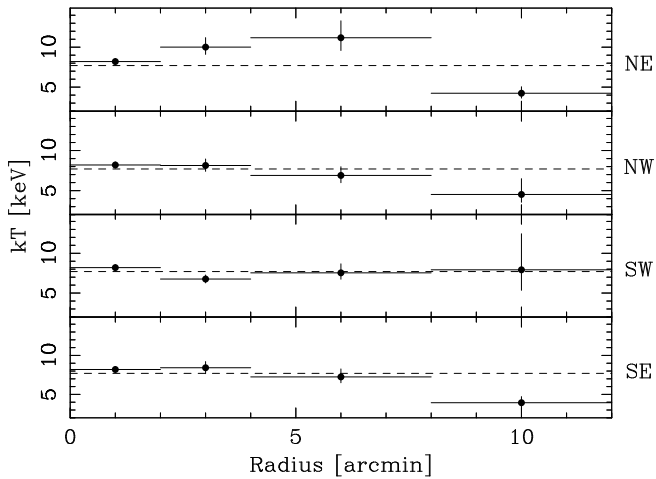,height=6.7cm,width=9cm,angle=-90}
}}
\vspace{-0.20in}
\caption{
Radial temperature profiles for the NE sector (first panel), the NW sector
(second panel), the SW sector (third panel) and the SE sector (forth
panel).
The temperature for the leftmost bin is derived from the entire circle,
rather than from each sector. The dashed lines indicate the average 
temperature derived from the radial profile presented in figure 1. 
}
\label{figmg7}
\end{figure}

\vspace{-0.15in}

In figure 3  we show  the temperature  
profiles obtained from the spectral fits for each of the 4 sectors.
In all the profiles we have included the temperature  
measure obtained for the central region with 
radius 2$^{\prime}$. 
Fitting each radial profile with a constant temperature we derive 
the following average sector temperatures: 
7.8$\pm$0.3 keV for the NE sector, 
8.0$\pm$0.3 keV for the NW sector, 
7.7$\pm$0.3 keV for the SW sector and 
7.5$\pm$0.3 keV for the SE sector. 
The fits yield the following $\chi^2$ values: 
$\chi^2 =26.45$ for 3 d.o.f. (Prob.$= 7.7\times 10^{-6}$) for the NE sector,
$\chi^2 =4.4$ for 3 d.o.f. (Prob.$= 0.22$) for the NW sector, 
$\chi^2 =5.0$ for 3 d.o.f. (Prob.$= 0.17$)  for the SW sector and 
$\chi^2 =25.4$ for 3 d.o.f. (Prob.$= 1.3\times 10^{-5}$) for the SE sector. 
In the NE sector  the temperature first increases to values \simgt 10 keV
in the second and third annulus, and then decreases to $\sim$ 5 keV in
the outermost annulus. 
By comparing the temperature of NE sector of the second and third annulus 
with the temperature averaged over the other 3 sectors  in the second 
and third annulus, we find that they differ at the
$\sim 2.5\sigma$ level.
In the SE  and NW sector the temperature decreases continuously as the 
distance from the cluster center increases, although the statistical
significance of the decrease is rather small in the NW sector. 
Finally in the SW sectors, due to the relatively large errors, no trend 
can be seen in the  temperature profile.

\begin{figure}
\vspace{-0.20in}
\centerline{
      \hbox{
      \psfig{figure=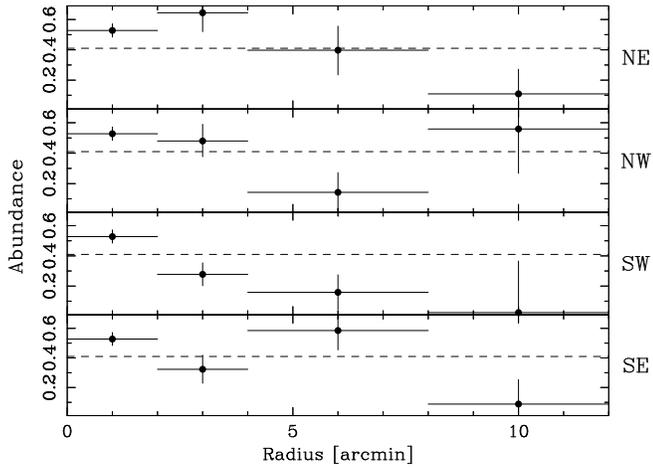,height=6.7cm,width=9.0cm,angle=-90}
}}
\vspace{-0.20in}
\caption{
Radial abundance profiles for the NE sector (first panel), the NW sector
(second panel), the SW sector (third panel) and the SE sector (forth
panel).
The abundance for the leftmost bin is derived from the entire circle,
rather than from each sector.
The dashed lines indicate the average 
abundance derived from the radial profile presented in figure 1. 
}
\label{figmg7}
\end{figure}

In figure 4 we show the abundance profiles for each of the 4 sectors.
In all profiles we have included the  
abundance measure obtained for the central region with 
bounding radius 2$^{\prime}$. 
Fitting each profile with a constant abundance we derive 
the following sector averaged abundances: 
0.50$\pm$0.04 for the NE sector,
0.49$\pm$0.04 for the NW sector, 
0.43$\pm$0.04 for the SW sector and  
0.48$\pm$0.04 for the SE sector.
The fits yield the following $\chi^2$ values: 
$\chi^2 =7.7$ for 3 d.o.f. (Prob.$= 5.3\times 10^{-2}$) for the NE sector,
$\chi^2 =8.0$ for 3 d.o.f. (Prob.$= 4.6\times 10^{-2}$) for the NW sector, 
$\chi^2 =15.7$ for 3 d.o.f. (Prob.$=1.3\times 10^{-3} $) for the SW sector and 
$\chi^2 =10.0$ for 3 d.o.f. (Prob.$= 1.8\times 10^{-2}$) for the SE sector. 
A decreasing trend is observed in all sectors, except perhaps the NW
sector. A highly statistically significant gradient is observed 
only in the SW and SE sectors.


\section{Discussion}

Previous measurements of the temperature structure of A2029
have been performed by S98 and White (1999), using ASCA data,
and by Irwin, Bregman \& Evrard (1999), using ROSAT PSPC data. 
S98  find a decreasing radial temperature 
profile. 
In figure 1 we have overlaid the temperature profile obtained by  
S98 using ASCA data, to our own BeppoSAX profile.
Although the single temperature measurements show some
discordance an overall temperature decline is observed in
both profiles.
Indeed, a fit with a linear profile of the type kT = a $ + $ b~r, 
where kT is in keV and r in arcminutes, to the S98 data, 
provides best fitting parameters:
a $= 9.6 \pm 1.3 $ keV, b$= -  0.35 \pm 0.28$ keV~arcmin$^{-1}$
compatible with those derived from the 
BeppoSAX data.
Recently White (1999) has reanalyzed the ASCA observation of
A2029 finding a temperature profile which, although suggestive of 
a temperature gradient, due to the rather 
large uncertainties, is consistent with a constant temperature and at the
same time with the BeppoSAX declining profile. 
Irwin, Bregman \& Evrard (1999) have used ROSAT PSPC hardness 
ratios to measure temperature gradients for a sample of nearby galaxy 
clusters, which includes A2029. 
In their analysis they find evidence of a radial temperature decrease,
the authors comment that a temperature gradient is probably 
present in this cluster.
The profiles we report in figure 3 suggest that the radial 
temperature gradient is most likely present in all sectors. 
We also find an indication of an azimuthal temperature gradient 
occurring in the annuli with bounding radii 2$^\prime$-4$^\prime$ 
(0.24 Mpc - 0.47 Mpc) and  4$^\prime$-8$^\prime$ (0.47 Mpc - 0.95 Mpc).
The data suggests that the NE sector of the cluster may be
somewhat hotter than the rest. Given the modest statistical significance
of this temperature enhancement, and the lack of detection 
of  substructure in this sector either from X-ray images (e.g. 
Buote \& Canizares 1996) or from optical velocity dispersion studies  
(Oegerle et al. 1995), we cannot make a strong case  
in favor of an azimuthal temperature gradient. 

The ASCA radial abundance profile reported
by S98 and by  White (1999) is characterized by rather large uncertainties, 
and is compatible with a constant abundance as well as with an 
abundance decrement such as the one we measure with BeppoSAX data.
The profiles we report in figure 4 suggest that the radial 
abundance gradient is most likely present in all sectors.


\begin{acknowledgements}
We acknowledge support from the BeppoSAX Science Data Center.
\end{acknowledgements}


\end{document}